\documentclass[aps,prl,preprint,superscriptaddress,showpacs,floatfix]{revtex4}
\usepackage{graphicx}
\usepackage{bm}
\begin{document}
\title{Exact solution of the five frequency model of the vacancy-assisted
impurity diffusion in the limit of vanishing vacancy concentration}
\author{V.~I.~Tokar}
\affiliation{
IPCMS-GEMM, UMR 7504 CNRS, 23, rue du Loess,
F-67034 Strasbourg Cedex, France
}
\affiliation{Institute of Magnetism, National Academy of Sciences,
36-b Vernadsky st., 03142 Kiev-142, Ukraine}
\date{\today}
\begin{abstract}
The van Hove autocorrelation function of the impurity is obtained which
is exact to the first order in the vacancy concentration.  It is found
that in the case of strong vacancy-impurity binding a singularity in
the van Hove function corresponding to a resonant bound state develops
on the unphysical sheet in the complex frequency plane close to the
real axis.  It is argued that this bound state corresponds to the
defect-impurity pairs widely used in models of diffusion in
semiconductors.
\end{abstract}
\maketitle
In close packed solids (like FCC crystals) the diffusion of atoms goes
predominantly via their exchange with vacancies.  Because the vacancy
concentration is usually small, in the limit of infinite dilution a
microscopic description of the diffusion reduces to the consideration of
the two-body interactions and exchanges between a single vacancy and
the atom of the diffusing specie \cite{philibert,manning,allnatt}.

When microscopic interactions are known, a two-body problem can usually
be solved exactly.  For example, in  Ref.\ \cite{tahir-kheli} a
solution for the tracer Green function accurate to the leading orders
in impurity and vacancy concentrations was obtained in the case of the
self-diffusion problem.  This success was due to the simplicity of the
self-diffusion case where all vacancy hopping parameters are
identical.

\begin{figure}
\begin{center}
\includegraphics[viewport = 0 20 229 206, scale = 0.45]{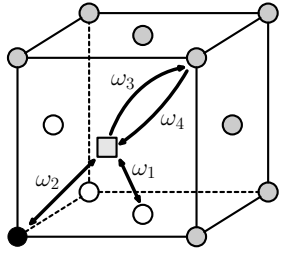}
\end{center}
\caption{\label{fig1}Frequencies of exchange of the vacancy (grey
square) with atoms in in the vicinity of the impurity (black circle) as
defined in the five-frequency model:  $\omega_1$--exchange with the
atoms in the first coordination sphere; $\omega_2$--exchange with the
impurity atom; $\omega_3$--dissociative jumps away from the impurity
into higher coordination spheres; $\omega_4$--associative jumps from
higher spheres into the first one.  Not shown in the figure is the
frequency $\omega_0$ of exchange with the atoms in the host bulk.}
\end{figure}

In the case of impurity, however, the vacancy in the host bulk and in
its vicinity experiences different forces, so the number of parameters
grows.  This case is most frequently described in the framework of the
so-called five-frequency model (5FM; see Fig.\ \ref{fig1})
\cite{philibert,manning,allnatt}.  Introduced by Lidiard \cite{
lidiard} half a century ago, this model has became a conventional tool
for the microscopic description of impurity diffusion.

The aim of the present Letter is to obtain the solution of the 5FM of
impurity diffusion similar to that of Ref.\ \cite{tahir-kheli} for the
tracer diffusion and to show that under strong vacancy-impurity(V-I)
binding the solution contains a resonant bound state which can be
identified with the defect-impurity pairs widely used in the
description of diffusion in semiconductors \cite{mulvaney,orlowski}.

To avoid awkward matrix formalism conventionally used in microscopic
theories of diffusion \cite{manning,allnatt}, the calculations of the
present Letter will be based on the second quantization representation
of stochastic dynamics \cite{doi}.  This allows for the treatment of
diffusion problems within the standard techniques of the quantum
theory.  A detailed exposition of this approach may be found in the
review article \cite{mattis}.  The formalism which will be used in this
Letter is presented in detail in Ref.\ \cite{me}, so below are listed
only the formulas used in subsequent calculations.

In the second quantization approach the host lattice completely filled
with the solvent atoms is considered to be the ground state
($|\text{host}\rangle$) of the diffusion problem.  The impurity atoms
and the vacancies are created (annihilated) on lattice site $i$ under
the action of the so-called hard-core boson creation (annihilation)
operators $a^+_i$ ($a_i$) and $b^+_i$ ($b_i$), respectively.  These
operators satisfy the canonical bosonic commutation relations in all
cases except when on the same site in which case they satisfy the
anticommutation relations \cite{grassberger}
\begin{equation} 
\label{ fermion}
\{d^+_i,d_i\}=1\quad\mbox{and}\quad\qquad d_i^2=(d^+_i)^2=0
\end{equation} 
(here $d_i$ stands either for $a_i$ or for $b_i$) to prevent multiple
occupancy of the site by the same specie.  The particle number
operators as usual are
\begin{equation} 
\label{ nv}
n_i = a^+_i a_i\mbox{\ \ and\ \ }v_i=b^+_i b_i
\end{equation} 
and the average concentrations are their statistical averages
$c_I=\langle n_i\rangle$ and $c_V=\langle v_i\rangle$.

The coherent state
\begin{equation} 
\label{ coherent}
|\rangle=\prod_i(1+a^+_i)(1+b^+_i)|\text{host}\rangle,
\end{equation} 
comprises all possible configurations of impurities and vacancies with
equal weight and is used below to compute statistical averages.  The
weak equivalences between the operators with respect to the conjugate
coherent state
\begin{equation} 
\label{ 2}
\langle|a_i = \langle|n_i;\qquad\langle|a^+_i = \langle|(1-n_i)
\end{equation} 
and similar relations for the vacancy operators are easily derived from
Eqs.\ (\ref{ fermion})--(\ref{ coherent}).

The matrix of transition probabilities ${\bm T}$ governing the time
evolution of a state vector
\begin{equation} 
\label{ evolution}
{|t\rangle} = \exp({\bm T}t)|0\rangle,
\end{equation} 
obeys the condition of probability conservation \cite{kadanoff}
\begin{equation} 
\label{ p=1}
\langle|{\bm T} = 0.
\end{equation} 

The van Hove autocorrelation function $G_{l0}$ describes the evolution
at equilibrium of an impurity atom initially placed at the origin of
the lattice coordinates.  It can be obtained as the $c_I\to0$ limit of
the impurity-impurity correlation function because in this limit only
the autocorrelations survive:
\begin{equation} 
\label{ G}
G_{l0}= \lim_{c_I\to0} 
\left[c_I^{-1}\langle|n_l\exp({t\bm T})n_0e^{-{\cal H}/{kT}}|\rangle\right],
\end{equation} 
where ${\cal H}$ is the Hamiltonian of interparticle interactions which
includes a normalization constant such that $\langle|e^{-{\cal
H}/{kT}}|\rangle=1$ \cite{me}.  With the use of Eq.\ (\ref{ p=1}) the
last equation can be re-written as a conventional retarded correlation
function.  For brevity, the following notation for such functions will
be used:
\begin{equation} 
\label{ AB}
\langle A;B\rangle_t\equiv \lim_{c_I\to0}
\left[c_I^{-1}\langle|A(t)B(0)e^{-{\cal H}/{kT}}|\rangle\right],
\end{equation} 
where $A(t)=e^{-t\bm T}Ae^{t\bm T}$.  Thus, the evolution of the
correlation function can be studied with the use of the 
Heisenberg equation 
\begin{equation} 
\label{ heisenberg}
\dot{A}=[A,{\bm T}]
\end{equation} 
provided the transition matrix is known.
According to Ref.\ \cite{kadanoff}, the transition matrix consists of
two parts:
\begin{equation} 
\label{ kadanoff}
{\bm T} = {\bm T}^{\text{in}} - {\bm T}^{\text{out}},
\end{equation} 
where the first part describes intersite particle exchange while the
second part is site-diagonal and depends only on the operators $n_i$
and $v_i$.  In the case under consideration the ``in'' part can be
written as
\begin{equation} 
\label{ T^in}
{\bm T}^{\text{in}}=\sum_{ij}\left(\omega_{ij}+
\omega_{ij}^xa^+_ja_i\right)b^+_ib_j.
\end{equation} 
Here the first term in the parenthesis corresponds to the vacancy
exchanges with the host atoms while the second term describes the V-I
exchanges.  In general case the hopping frequencies entering
Eq.\ (\ref{ T^in}) depend on the whole many-body configuration of
vacancies and impurities.  But in the double dilute limit under
consideration it is sufficient to keep the dependence of $\omega_{ij}$
on the position of only one impurity, while $\omega_{ij}^x$ does not
depend on the configuration at all because the two participating
particles are already present in the transition matrix.  So in the
two-body approximation $\omega_{ij}^x$ can be identified with
$\omega_2$ while $\omega_{ij}$ have the values shown in
Fig.\ \ref{fig1}.  It is important to note that
$\omega_{ij}\not=\omega_{ji}$.

The ``out'' term in Eq.\ (\ref{ kadanoff}) can be easily obtained from
Eq.\ (\ref{ p=1}) with the use of Eq.\ (\ref{ 2}).  But because it
commutes with all site-diagonal operators, it will not be needed in the
calculations below.  Thus, substituting $A = n_i$ and ${\bm
T}^{\text{in}}$ into Eq.\ (\ref{ heisenberg}) and computing the average
in Eq.\ (\ref{ AB}) with $B=n_0$ to the leading orders in $c_I$ and
$c_V$, with the use of Eq.\ (\ref{ 2}) it is straightforward to obtain
\begin{equation} 
\label{ cT}
d G_{l0}(t)/d t =
\omega_{2}\sum_{\bm{\delta}}\left(\langle 
n_{l+{\bm{\delta}}}v_l;n_{0}\rangle_t
-\langle n_lv_{l+{\bm{\delta}}};n_{0}\rangle_t
\right),
\end{equation} 
where by $\bm{\delta}$ are denoted 12 different lattice 
vectors connecting nearest neighbor sites.

Similarly, by choosing $A=n_lv_j$ and $B=n_0$ in Eqs.\ (\ref{ AB}) and
(\ref{ heisenberg}) one arrives at a closed set of equations necessary
to calculate the impurity van Hove function $G_{l0}(t)$ to the chosen
order of approximation
\begin{eqnarray} 
\label{ eq1}
&&(d/d t) \langle n_lv_j;n_{0}\rangle_t =
 \omega_{jl}^x\langle n_jv_l;n_{0}\rangle_t- 
\omega_{lj}^x\langle n_lv_j;n_{0}\rangle_t
\nonumber\\
&&+\sum_i\omega_{ji}\langle n_l 
v_i;n_{0}\rangle_t-(\sum_{i}\omega_{ij})
\langle n_l v_j;n_{0}\rangle_t.
\end{eqnarray} 
The above two equations have the same formal structure as the
corresponding equations of Ref.\ \cite{tahir-kheli}, so their solution
can be found by a straightforward use of the techniques of that
reference.  Therefore, only notation and the points of departure from
the solution of Ref.\ \cite{tahir-kheli} will be explained in some
detail.

The first step is to subject the equations to the Laplace (with respect
to $t$) and spatial Fourier transforms (L-F transform).  Their
normalizations should be understood from the definition of $f_{\bm
K}({\bm r})$ (cf.\ Ref.\ \cite{tahir-kheli})
\begin{equation} 
\label{ G2_def}
\langle n_lv_{l-r};n_{0}\rangle_t
=\frac{1}{2\pi i}\oint dz\frac{1}{N}\sum_{\bm K}
f_{\bm K}({\bm r})e^{zt+i{\bm K}\cdot l}
\end{equation} 
and a similar expression for $G_{l0}(t)$-$G_{\bm K}(z)$ pair.  The
Laplace transform of derivatives contains the function values at $t=0$.
In the present case they are given by the equilibrium correlators in
(\ref{ AB}) which are $G_{0l}(0)=\delta_{l0}$ and
\begin{equation} 
\label{ nvn0}
\langle n_l v_jn_{0}\rangle_0=c_V\delta_{l0}(1-\delta_{lj})[1+(e^{E_b/kT}-1),
\delta_{|l-j|,|{\bm \delta|}}],
\end{equation} 
where the multiplier in the parentheses forbids simultaneous occupation
of the same site by the impurity and the vacancy while the term in
square brackets describes the enhancement (suppression) of the vacancy
concentration in the first coordination sphere of the impurity provided
the V-I binding energy $E_b$ is positive (negative).  The transformed
Eq.\ (\ref{ cT}) reads (cf.\ Eq.\ (5) of Ref. \cite{tahir-kheli})
\begin{equation} 
\label{ first}
z G_{\bm K}(z) = 1 -
\omega_{2}\sum_{\bm{\delta}}[1-\exp(i{\bm K}
\cdot\bm{\delta})]f_{\bm K}(\bm{\delta}).
\end{equation} 
According to Ref.\ \cite{tahir-kheli} it should be cast into the form of 
the diffusion propagator
\begin{equation} 
\label{ dyson}
G_{\bm K}(z) = \left[z+c_VF(z,{\bm K})\epsilon_{\bm K}\right]^{-1},
\end{equation} 
with the $z$ and ${\bm K}$-dependent diffusion correlation factor
\begin{equation} 
\label{ F}
F(z,{\bm K}) = \frac{\omega_2z}
{c_V\epsilon_{\bm K}}\sum_{\bm{\delta}}[1-\exp(i{\bm
K}\cdot\bm{\delta})]f_{\bm K}(\bm{\delta}),
\end{equation} 
where $ \epsilon_{\bm K} = \omega_0\sum_{{\bm\delta}}\left[1-\exp({i{\bm 
K}\cdot{\bm\delta}})\right]$. 

Eq.\ (\ref{ eq1}) under the L-F transform takes the form
\begin{eqnarray} 
\label{ theEQ}
&&zf_{\bm K}({\bm r})=c_V[1-\delta({\bm r})]
[1+(e^{\frac{E_b}{kT}}-1)\sum_{\bm{\delta}}\delta({\bm r}
-\bm{\delta})]\nonumber\\
&&+\omega_2\sum_{\bm{\delta}}\delta({\bm r} 
-\bm{\delta})[\exp(-i{\bm K}\cdot\bm{
\delta})f_{\bm K}(-{\bm r}) - f_{\bm K}({\bm r})]\nonumber\\
&&+\sum_{\bm{\delta}}\left[\omega({\bm r} 
- \bm{\delta},\bm{\delta})f_{\bm K}({\bm r}-\bm{\delta})
-\omega({\bm r},\bm{\delta})f_{\bm K}({\bm r})\right],
\end{eqnarray} 
where $\omega({\bm r},\bm{\delta})$ is the frequency of hopping from
site ${\bm r}$ to ${\bm r}+\bm{\delta}$.  It is easy to see that
$f_{\bm K}(0)=0$ satisfies this equation irrespective of $f_{\bm
K}({\bm r})$ values at other sites because in the 5FM the vacancy hops
onto or from the site occupied by the impurity are forbidden:
$\omega(- \bm{\delta},\bm{\delta})=\omega(0,\bm{\delta})=0$.
Therefore, henceforth $f_{\bm K}(0)$ will be excluded from the
consideration.  Furthermore, because in the absence of fluxes (${\bm
K}=0)$ the system is at equilibrium, the values of $f_{\bm K=0}({\bm
r})$ should be equal to the L-F transformed Eq.\ (\ref{ nvn0}).  The
result can be written symbolically as
\begin{equation} 
\label{ f_0}
\vec{f}_{0} = \vec{E}/z,
\end{equation} 
where the vector components are the values of corresponding quantities
at lattice sites ${\bm r}$.  Because the components of $\vec{E}$
coincide with the first line on the right hand side (RHS) of
Eq.\ (\ref{ theEQ}), $z\vec{f}_{0}$ turns this line into identity.
Thus, in order to satisfy the rest of equation the frequencies should
be subject to the constraint \cite{philibert,manning,allnatt}
\begin{equation} 
\label{ epsilon}
{\omega_3}/{\omega_4}=\exp(-E_b/kT)\equiv\varepsilon
\end{equation} 
as can be checked by a direct substitution of $\vec{f}_{0}$ from
Eq.\ (\ref{ f_0}) into the last line of Eq.\ (\ref{ theEQ}).

The method of solution for general ${\bm K}$ proposed in
Ref.\ \cite{tahir-kheli} essentially consists in subtraction from both
sides of Eq.\ (\ref{ theEQ}) of the term which formally coincides with
the equation last line but with all $\omega({\bm r},{\bm r}^{'})$
replaced with $\omega_0$.  The result can be symbolically written as
\begin{equation} 
\label{ symb}
\hat{R}^{-1}\vec{f}_{\bm K} = \vec{E} + \hat{V}\vec{f}_{\bm K},
\end{equation} 
where matrix $\hat{R}$ is composed of matrix elements
\begin{equation} 
\label{ R}
R_{{\bm r},{\bm r}^{'}} = \frac{1}{N}\sum_{\bm \lambda}
\frac{\exp[-i{\bm \lambda}\cdot({\bm r}-{\bm r}^{'})]}
{z+\epsilon_{\bm \lambda}},
\end{equation} 
the first line of Eq.\ (\ref{ theEQ}) is represented by $\vec{E}$ and
the remaining terms are gathered into $\hat{V}\vec{f}_{\bm K}$.  The
meaning of this transformation is that now on the RHS of Eq.\ (\ref{
symb}) only a finite number of $f_{\bm K}({\bm r})$ (54 in this case)
with ${\bm r}$ belonging to the first four coordination spheres remain
because beyond these spheres $\omega({\bm
r},\bm{\delta})-\omega_0\equiv0$, as can be seen from the definition of
frequencies in Fig.\ \ref{fig1}.  Now multiplying both sides of
Eq.\ (\ref{ symb}) with $\hat{R}$
\begin{equation} 
\label{ theEQ1}
\vec{f}_{\bm K} = \hat{R}\vec{E} + \hat{R}\hat{V}\vec{f}_{\bm K}
\end{equation} 
and retaining only those equations which have at their left hand sides
the same $f_{\bm K}({\bm r})$ that are present on the RHS one obtains a
linear system of 54 equations which can be solved, e.\ g.\, with the
use of the Cramer's rule as
\begin{equation} 
\label{ cramer} 
f_{\bm K}({\bm r}) = \Delta_{\bm r}(z,{\bm K})/\Delta(z,{\bm K}).
\end{equation} 
Here $\Delta(z,{\bm K})$ is the determinant of the system and
$\Delta_{\bm r}(z,{\bm K})$ is the same determinant with the column
corresponding to $f_{\bm K}({\bm r})$ replaced by $\hat{R}\vec{E}$.
Substituting $f_{\bm K}({\bm \delta})$ thus obtained into Eqs.\ (\ref{
F}) and (\ref{ G}) one obtains $G_{\bm K}(z)$ for general values of $z$
and ${\bm K}$.  It can be used, e.\ g., to study the diffusion of
M\"ossbauer impurities where the knowledge of the van Hove
autocorrelation function at finite values of $z$ and ${\bm K}$ is
essential \cite{voglAlFe}.

If one is interested only in the diffusion limit $z, {\bm K}\to 0$,
then the size of the system (\ref{ theEQ1}) can be reduced to 13 by the
choice of a high symmetry direction ${\bm K} = (K,0,0)$
\cite{tahir-kheli}.  It can be shown that the correlation factor
$F(0,0)$ in Eq.\ (\ref{ F}) can be expressed through the ratio of two
determinants of size 13.  The corresponding expression was derived and
numerically checked on thousands of randomly generated frequency
quintets.  In all cases excellent agreement with the approximate
expression due to Manning \cite{manning} was found.  In particular, it
was numerically confirmed that in the case of strong V-I binding the
correlation factor is always enhanced as $O(1/\omega_3)$.  This might
have been guessed on the basis of Eq.\ (\ref{ f_0}) which states that
$f_{\bm K=0}({\bm \delta})=\varepsilon^{-1}/z\propto 1/\omega_3$.  A
careful analysis shows that in Eq.\ (\ref{ cramer}) the $\omega_3^{-1}$
factor can originate only from the denominator because in the numerator
the contribution due to $\hat{R}\vec{E}$ containing $\varepsilon^{-1}$
is suppressed in the limit $z\to0$ by the factor z in Eq.\ (\ref{ F})
(see Eq.\ (\ref{ F_approx}) below).  Thus, in the case of strong V-I
binding the determinant $\Delta(0,0)$ should be proportional to
$\omega_3$.
Then it is reasonable to assume that at small $z$ and $|{\bm K}|$
\begin{equation} 
\label{ det}
\Delta(z,{\bm K})\approx C_1\varepsilon+C_2z +C_3K^2,
\end{equation} 
where $C_k$ are some constants.  From here it would follow that
$F(z,{\bm K})$ develop a pole-like singularity at some small value of
$z$.

To qualitatively assess the influence of this singularity on the
impurity diffusion let us consider the simplest model with V-I binding:
the so-called 2-frequency model (2FM) \cite{philibert,manning,allnatt}
where all frequencies are equal to $\omega_0$ except $\omega_3$ which
according to Eq.\ (\ref{ epsilon}) should be equal to
$\varepsilon\omega_0$.  In this case the dimension of determinants
reduces to 3 so with the use of exact relations between $R_{{\bm r},{\bm
r}^{'}}$ derived in Ref.\ \cite{GreenFCC} one arrives after some
algebra to an expression which to leading orders in the small
quantities reads
\begin{equation} 
\label{ F_approx}
F_{2FM}(z,{\bm K}) \approx \frac{g+0.5z\varepsilon^{-1}}
{z+2g\varepsilon+(\omega_0/6)(aK)^2},
\end{equation} 
where $g=(24R_{0,\bm{\delta}})^{-1}$.  Because $R_{0,\bm{\delta}}$ has
a square root singularity at $z=0$ \cite{GreenFCC}, the pole shifts
from the real axis on unphysical sheet on the distance
$O(\varepsilon^{3/2})\ll 1$, as can be seen from the denominator of
Eq.\ (\ref{ F_approx}).

\begin{figure}
\begin{center}
\includegraphics[viewport = 0 10 386 297, scale = 0.5]{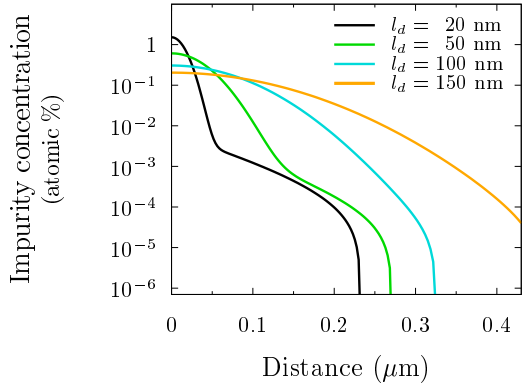}
\end{center}
\caption{\label{fig2}(color online) One-dimensional diffusion 
along (1,0,0) direction of the FCC lattice of initial $\delta$-function
impurity distribution calculated in the two frequency model
(see the text) for different values of diffusion length 
$l_d = 2(D_It)^{1/2}$.  For definiteness, the parameters 
corresponding to the arsenic impurity in silicon at temperature 
$T=800{\,}^\circ$C were used: $E_b\approx1.2$ eV \cite{AsV2}, 
$c_V = 10^{-8}$ \cite{c_vSi}, and $a = 5.43$~\AA.}
\end{figure}

The quasiparticle content of the van Hove function can be established
by the substitution of $F_{2FM}$ into Eq.\ (\ref{ F}) and by
approximating $g$ with a real constant $g_0=g(z=0)$.  In this case the
van Hove function acquires simple two-pole structure
\begin{equation} 
\label{ approx}
G_K(z) \approx \frac{1-\phi(K)}{z+z_1(K)}+\frac{\phi(K)}{z+z_2(K)},
\end{equation} 
where
\begin{eqnarray} 
\label{ z1}
&&z_1(K)\approx \frac{c_v}{2\varepsilon}\omega_0(aK)^2\equiv
D_IK^2,\\
\label{ z2}
&&z_2(K)\approx 2g_0\varepsilon+(\omega_0/6)(aK)^2,
\end{eqnarray} 
and
\begin{equation} 
\label{ phi}
\phi(K) = ({c_v}/{12\varepsilon})\omega_0^2(aK)^4/z_2^2(K).
\end{equation} 
Thus, there are two diffusion poles. The one at $-z_1$ corresponds to
the interaction-enhanced impurity diffusion and the pole at $-z_2$ can
be identified with the V-I pair state.  Indeed, being the part of the
impurity autocorrelation function this state obviously contains an
impurity.  On the other hand, its diffusivity is independent of $c_v$
which means that the vacancy is always present in this state.  This is
further confirmed by the diffusion profiles computed as inverse L-F
transform of Eq.\ (\ref{ approx}) and shown in Fig.\ \ref{fig2}.  At
small diffusion lengths they exhibit characteristic tails which in pair
diffusion models \cite{mulvaney,orlowski} are attributed to the
diffusing defect-impurity pairs.

Unfortunately, these profiles are not quite physical because at large
distances they acquire small negative values.  More accurate
calculation of the inverse L-F transform is needed to deside on whether
this deficiency is because of too crude approximations made above or
that $O(c_v)$ approximation is insufficient at large distances.
\begin{acknowledgments} I am indebted to NATO for supporting my one
year stay at IPCMS where the major part of this work was done.  I
express my gratitude to University Louis Pasteur de Strasbourg and
IPCMS for their hospitality and to Hugues Dreyss\'e for collaboration.
I am grateful to R\'en\'e Monnier for making possible my visit to ETH
(Z\"urich) where this work was started.  I would like to thank Jarek
D\c{a}browski for attracting my attention to the pair diffusion,
Camilla Schmidt for useful bibliographical references, and
Prof.\ A.\ B.\ Lidiard for interest in the work.
\end{acknowledgments}
\bibliographystyle{apsrev}

\end{document}